\documentclass[letterpaper]{iopart}
\usepackage{iopams}
\usepackage{graphicx}
\usepackage{amssymb}
\usepackage{latexsym}
\usepackage[dvips]{epsfig}

\newcommand{\derv}[2]{ \frac{d #1}{d #2}}

\newcommand{\be}{\begin{equation}}
\newcommand{\ee}{\end{equation}}
\newcommand{\bea}{\begin{eqnarray*}}
\newcommand{\eea}{\end{eqnarray*}}
\newcommand{\bean}{\begin{eqnarray}}
\newcommand{\eean}{\end{eqnarray}}
\newcommand{\overleftrightarrow}[1]{\vbox{\ialign{##\crcr
    $\leftrightarrow$\crcr\noalign{\kern-1pt\nointerlineskip}
    $\hfil\displaystyle{#1}\hfil$\crcr}}}

\newcommand{\n}[1]{\label{#1}}

\newcommand{\mbq}{\mathbf{q}}
\newcommand{\mbf}{\mathbf{f}}

\begin{document}
\title{Critical Collapse of an Ultrarelativistic Fluid in
  the $\Gamma\to 1$ Limit}
\author{Martin Snajdr}
\address{Department of Physics and Astronomy,
     University of British Columbia,
     Vancouver BC, V6T 1Z1 Canada}

\begin{abstract}      
In this paper we investigate the critical collapse of an ultrarelativistic 
perfect fluid with the equation of state $P=(\Gamma-1)\rho$ in the limit of
$\Gamma\to 1$. We calculate the limiting continuously self similar (CSS) solution and
the limiting scaling exponent by exploiting self-similarity of the solution.
We also solve the complete set of equations governing the gravitational collapse
numerically for $(\Gamma-1) = 10^{-2},\dots,10^{-6}$ and compare them with
the CSS solutions. We also investigate the supercritical regime and discuss the
hypothesis of naked singularity formation in a generic gravitational collapse.
The numerical calculations make use of advanced methods such as high 
resolution shock capturing evolution scheme for the matter evolution, adaptive
mesh refinement, and quadruple precision arithmetic. The treatment of vacuum is 
also non standard. We were able to tune the 
critical parameter up to $30$ significant digits and to calculate the scaling exponents
accurately. The numerical results agree very well with those calculated using 
the CSS ansatz. The analysis of the collapse in the supercritical regime supports the 
hypothesis of the existence of naked singularities formed during a generic gravitational collapse.
\end{abstract}

\section{Introduction}
\setcounter{equation}0
Since the discovery of critical phenomena in gravitational collapse of
a massless scalar field by Choptuik in 1993 \cite{Choptuik93}
researchers have observed critical phenomena in many different matter models.
In this paper we focus on perfect fluid, in particular the 
ultrarelativistic perfect fluid with the equation of state 
\be
  P = (\Gamma-1)\rho\ .
\n{eq1.1}
\ee
The study of critical phenomena in ultrarelativistic perfect fluids has quite 
interesting history.
It started with Evans and Coleman's calculations \cite{Evans_Coleman} who were able to
obtain critical solution for the radiation fluid ($\Gamma=4/3$) by employing 
the CSS ansatz and also by directly solving the PDE's.
Later, Koike \etal \cite{Koike_Hara_Adachi95} used perturbation analysis to calculate the 
scaling exponent of the radiation fluid collapse.
Maison \cite{Maison} generalized the calculations for $1.01<\Gamma<1.888$ and similar analysis
was carried out by Hara, Koike and Adachi \cite{Hara_Koike_Adachi96}.
Both speculated that the solutions might not exist for $\Gamma>1.89$.
Fully numerical treatment of the collapse for $1.05<\Gamma<1.5$ was performed
by Evans and Perkins~\cite{Perkins}.
The inability to obtain the solutions for $\Gamma>1.89$ lead to various 
ideas about what happens in that regime. Some proposed that the solution 
changes from CSS into DSS (discretely self similar) or a mixture of both.
It was also proposed that the critical solution might be of type I and not
type II.

The confusion was definitely resolved by Neilsen and Choptuik \cite{Neilsen_Choptuik2000} who
demonstrated by directly solving the PDE's that nothing special happens
for $\Gamma>1.89$. They carried out the calculations all the way up to 
$\Gamma=2$. Moreover, they were able to obtain the solutions for $\Gamma>1.89$ 
with the CSS ansatz. The key to success was to use arbitrary precision arithmetic.

The purpose of this paper is several. We confirm the results of the 
perturbation calculations
with the CSS ansatz from previous works for values of $\Gamma$ close to $1$ and,
more importantly, we confirm by direct numerical collapse calculations that the
 perturbative
calculations give the correct solutions for that regime.
We obtain the exact numerical value of the limiting scaling exponent that
has never been calculated before. 

One of the reasons for studying the $\Gamma\to 1$ limit is to verify the hypothesis
made by Harada and Maeda that for $\Gamma<1.0105$ the final state of a
supercritical collapse is not a black hole but a naked singularity 
\cite{Harada_98,Harada_Maeda_2001}.
It is known that the critical solution contains a naked singularity but 
this is only achieved by fine tuning of the critical parameter.
The naked singularity formed during the supercritical collapse would not
require any fine tuning and therefore is much more interesting.

Critical collapse possesses a unique set of challenges for numerical treatment.
During the evolution the dynamics occurs on ever decreasing length scales and the
magnitudes of the fluid state variables increase by many orders. 
Therefore, advanced numerical techniques have to be used to perform these calculations.

For this purpose we develop a general adaptive mesh refinement (AMR) algorithm which is
capable of performing the critical collapse calculations.
This algorithm is general and does not rely on prior knowledge of the dynamics.
We also use innovative treatment of vacuum that does not use a low density artificial
atmosphere --- a method widely used to ameliorate numerical instabilities at the
matter--vacuum boundary.

The use of quadruple precision arithmetic and the above-described advanced numerical 
techniques allow us to obtain the solutions and scaling exponents with higher accuracy
than in previous works.

The organization of the paper is as follows.
Section~\ref{sec:hydro} reviews the basic equations of
general relativistic hydrodynamics in spherical symmetry.
In section~\ref{sec:CSS} we sketch the ideas behind the CSS ansatz and
present the equations and their solutions for the limiting case $\Gamma\to 1$.
Similarly, in section~\ref{sec:perturb} we briefly review the perturbation theory
and present the equations and their solutions for the relevant eigenmodes in the
limiting case.
In the later two sections we closely follow the approach described in \cite{Hara_Koike_Adachi96}, 
 \cite{Koike_Hara_Adachi99} where all of the missing details can be found.
Section~\ref{sec:results} describes the method of obtaining the critical scaling exponent
from numerical calculations and a way of estimating the errors.
It summarizes and compares results obtained from both approaches.
We also compare our numerical results from supercritical calculations with
the expected CSS solutions.
In section~\ref{sec:numerics} we describe in more details some of the features of the
numerical algorithms such as AMR and vacuum treatment.
Section~\ref{sec:conclusion} summarizes the results and concludes.

Throughout this paper we use geometrized units with $G=c=1$.

\section{Hydrodynamics in spherical symmetry}
\label{sec:hydro}
\setcounter{equation}0

In our calculations we use spherical polar coordinates in which the
metric has the form
\be
  ds^2 = -\alpha(t,r)^2 dt^2 + a(t,r)^2 dr^2 + r^2 d\Omega^2\ .
\n{eq2.1}
\ee
It is convenient to define an auxiliary variable $m(t,r)$
\be
  m(t,r) = \frac{r}{2}\left(1-\frac{1}{a(t,r)^2}\right)\ ,
\n{eq2.2}
\ee
which could be interpreted as the total mass enclosed within the radius $r$.
In the vacuum regions we demand that the metric reduces to the Schwarzschild
one, i.e., 
\be
  a(t,r) = -\frac{1}{\alpha(t,r)}\ .
\n{eq2.3}
\ee 

The stress-energy tensor for perfect fluid has the form
\be
  T^{\mu\nu} = (\rho+P) u^\mu u^\nu + P g^{\mu\nu}\ ,
\n{eq2.4}
\ee
where $P$ is the isotropic pressure, $u^\mu$ is the $4$-velocity
of the fluid element and $\rho$ is the {\em total energy density} as seen by
Eulerian observers (see later).
The energy density $\rho$ can be subdivided into two contributions
\be
  \rho = \rho_0 (1+\epsilon)\ ,
\n{eq2.5}
\ee
with $\rho_0$ being the {\em rest mass density} and $\epsilon$ being the 
{\em specific internal energy}.
In the ultrarelativistic limit the internal energy contribution dominates
the total energy density, i.e., 
\be
  \epsilon \gg 1\ .
\n{eq2.6}
\ee

The Eulerian observers' $4$-velocity $u^\mu=n^\mu$, where
$n^\mu$ is the unit normal vector to the spatial hypersurfaces defined by the constant
time $t$ slices.
These observers see the fluid moving with the $3$-velocity 
\be
  v^r = \frac{u^r}{\alpha u^t}\ .
\n{eq2.7}
\ee
It is useful to define a velocity $v$ as
\be
  v = a v^r
\n{eq2.8}
\ee
since the Lorentz factor is then a simple function of $v$
\be
  W = \alpha u^t = \frac{1}{\sqrt{1-v^2}}\ .
\n{eq2.9}
\ee

The equations of motion for the fluid can be derived from 
conservation laws
\be
  T^{\mu\nu}_{\ \ \ ;\nu}=0\ ,
\n{eq2.10}
\ee
\be
  J^{\mu}_{\ \ ;\mu}=0\ ,
\n{eq2.11}
\ee
where 
\be
  J^\mu = \rho_0 u^\mu\ .
\n{eq2.12}
\ee

These equations must be supplemented with an equation of state
which for perfect fluid has the form
\be
  P = (\Gamma-1) \rho_0\epsilon\ .
\n{eq2.13}
\ee
In the ultrarelativistic limit described by \eref{eq2.6}
the equation of state has the form
\be
  P = (\Gamma-1)\rho\ .
\n{eq2.14}
\ee
The equation of state \eref{eq2.14} is the one we use.
The set of {\em primitive variables}  $(\rho,v)$ characterize the state of the fluid completely.
Note that since $\rho_0$ does not enter the equation of state \eref{eq2.14}
the equation \eref{eq2.11} (baryon number conservation) is no longer
needed for the dynamical evolution of ultrarelativistic fluid.

We define a complementary set of {\em conservative} fluid variables  $(S,E)$
\be
  S = (\rho +P) W^2\, v\ ,
\n{eq2.15}
\ee
\be
  E = (\rho +P) W^2 - P\ .
\n{eq2.16}
\ee
The equations of motion for the fluid have the form
\be
  \dot{S} + \frac{1}{r^2}\left[r^2 X (Sv + P)\right]' = \Psi ,
\n{eq2.17}
\ee
\be
  \dot{E} + \frac{1}{r^2}\left[r^2 X S\right]' = 0\ ,
\n{eq2.18}
\ee
where 
\be
  X=\frac{\alpha}{a}
\n{eq2.19}
\ee
and the source term $\Psi$ has the form
\be
  \Psi= (S v-E)\left(8\pi\alpha a r P + \alpha a \frac{m}{r^2}\right)+\alpha a P\frac{m}{r^2}
           + \frac{2\alpha P}{a r}.
\n{eq2.20}
\ee
The dot denotes differentiation with respect to the coordinate time whereas prime denotes
differentiation with respect to the radial coordinate $r$.

As has been argued in \cite{Neilsen_Choptuik2000} in numerical applications 
it is convenient to work with a derived conservative quantities
\be
  \Phi = E-S\ ,
\n{eq2.21}
\ee
\be
  \Pi = E+S\ .
\n{eq2.22}
\ee
The equations of motion for these new variables have the form
\be
  \dot{\mbq} + \frac{1}{r^2}\left[r^2 X \mbf \right]' = \mathbf{\Sigma} , 
\n{eq2.23}
\ee
with
\be
\fl
  \mbq = \left(\begin{array}{c}
           \Pi\\
           \Phi
         \end{array}\right)\ ,\hspace{1cm}
  \mbf = \left(\begin{array}{c}
           \frac{1}{2}(\Pi-\Phi)(1+v)+P\\
           \frac{1}{2}(\Pi-\Phi)(1-v)-P
         \end{array}\right)\ ,\hspace{1cm}
  \mathbf{\Sigma} = \left(\begin{array}{c}
                      \Psi\\
                      -\Psi
                    \end{array}\right)\ .
\n{eq2.24}
\ee

The equations for the geometry can be obtained from the non-trivial component
of the momentum constraint
\be
  \dot{a} = -4\pi r\alpha a^2 S\ ,
\n{eq2.25}
\ee
the polar slicing condition 
\be
  \frac{\alpha'}{\alpha} = a^2\left(4\pi r(Sv+P) + \frac{m}{r^2}\right)\ ,
\n{eq2.26}
\ee
and the Hamiltonian constraint
\be
  \frac{a'}{a} = a^2\left(4\pi r E-\frac{m}{r^2}\right)\ .
\n{eq2.27}
\ee

In the numerical calculations we use only equations 
(\ref{eq2.26}) and (\ref{eq2.27}), i.e., we use a fully constrained evolution for the geometry.
\section{CSS solutions}
\label{sec:CSS}
\setcounter{equation}0

Under the assumption that the critical solutions are continuously self similar we can 
transform the system of PDE's (\ref{eq2.23}) and (\ref{eq2.25})--(\ref{eq2.27})
 into system of ODE's that can then be solved numerically.
Once the critical solution is found it is possible to solve for linear perturbation modes.
The {\em relevant mode} is the mode that has the eigenvalue with the largest real part.
There exists a simple relation between the eigenvalue of the relevant mode and
the scaling exponent
\be
  \gamma = \frac{1}{\kappa}\ ,
\n{eq3.1}
\ee
where $\kappa$ is the eigenvalue of the relevant mode (in our case it is a 
real number).

Let us now briefly summarize the procedure of finding the CSS solutions.
The details can be found in \cite{Hara_Koike_Adachi96}.
We can rewrite the equations (\ref{eq2.23}), (\ref{eq2.25})--(\ref{eq2.27}) in terms of new variables adapted to the CSS symmetry
\be
  s = -\ln(-t_*)\ ,
\n{eq3.2}
\ee
\be
  x = \ln(-\frac{r}{t_*})\ .
\n{eq3.3}
\ee
The time coordinate $t_*$ is fixed by the requirement that the collapsing 
CSS solution reaches the origin at time $t_* = 0$ and that the sonic point
is always located at $x=0$.
Using a new set of variables 
\be
  N = \frac{\alpha}{ae^x}\ ,
\n{eq3.4}
\ee
\be
  A = a^2\ ,
\n{eq3.5}
\ee
\be
  \omega = 4\pi r^2 a^2 \rho\ ,
\n{eq3.6}
\ee
we can write the system
\be
  \frac{A_{,x}}{A} = 1-A+\frac{2\omega (1+(\Gamma-1) v^2}{1-v^2}\ ,
\n{eq3.7}
\ee
\be
  \frac{N_{,x}}{N} = -2 + A - (2-\Gamma)\omega\ ,
\n{eq3.8}
\ee
\be
  \frac{A_{,s}}{A} + \frac{A_{,x}}{A} = - \frac{2\Gamma N v \omega}{1-v^2}\ ,
\n{eq3.9}
\ee
\[
\fl
  \frac{(\Gamma - 1) v}{\omega}\omega_{,s} + \frac{(1+N v)}{\omega}\omega_{,x}
  + \frac{\Gamma (N+v)}{1-v^2}v_{,x} 
\]
\be
= \frac{3(2-\Gamma)}{2}Nv - \frac{2+\Gamma}{2}A N v
 + (2-\Gamma)Nv\omega\ ,
\n{eq3.10}
\ee
\[
\fl
  \frac{(\Gamma - 1) v}{\omega}\omega_{,s} + \frac{\Gamma}{1-v^2}v_{,s} + 
  (\Gamma-1)\frac{N+v}{\omega}\omega_{,x} + \frac{\Gamma(1+Nv)}{1-v^2}v_{,x}
\]
\be
 = (2-\Gamma)(\Gamma-1)N\omega \frac{7\Gamma-6}{2}N + \frac{2-3\Gamma}{2}A N\ .
\n{eq3.11}
\ee
These equations are not independent --- the equation \eref{eq3.9} is automatically 
satisfied if the rest of them are providing the solutions satisfy the boundary conditions
\[
  A(s,-\infty) = 1\ ,
\]
\be
  v(s,-\infty) = \omega(s,-\infty) = 0\ . 
\n{eq3.12}
\ee

For a CSS solution all of the variables $(N,A,v,\omega)$ are functions of $x$ only
therefore the equations \eref{eq3.7}--\eref{eq3.11} reduce to a set of
ODE's
\be
  \frac{A_{,x}}{A} = 1-A+\frac{2\omega (1+(\Gamma-1) v^2}{1-v^2}\ ,
\n{eq3.13}
\ee
\be
  \frac{N_{,x}}{N} = -2 + A - (2-\Gamma)\omega\ ,
\n{eq3.14}
\ee
\be
  \frac{A_{,x}}{A} = - \frac{2\Gamma N v \omega}{1-v^2}\ ,
\n{eq3.15}
\ee
\be
\fl
  \frac{(1+N v)}{\omega}\omega_{,x} + \frac{\Gamma (N+v)}{1-v^2}v_{,x} = 
  \frac{3(2-\Gamma)}{2}Nv - \frac{2+\Gamma}{2}A N v + (2-\Gamma)Nv\omega\ ,
\n{eq3.16}
\ee
\be
\fl
  (\Gamma-1)\frac{N+v}{\omega}\omega_{,x} + \frac{\Gamma(1+Nv)}{1-v^2}v_{,x}
 = (2-\Gamma)(\Gamma-1)N\omega \frac{7\Gamma-6}{2}N + \frac{2-3\Gamma}{2}A N\ .
\n{eq3.17}
\ee
From (\ref{eq3.13}) and (\ref{eq3.15}) an algebraic equation can be formed
\be
  (1-A)(1-v^2)+2\omega (1+(\Gamma-1) v^2 = - 2\Gamma N v \omega\ ,
\n{eq3.18}
\ee
which can be, in principle, used to eliminate one of the variables.
We will label the CSS solutions as $(N_{\rm ss},A_{\rm ss},v_{\rm ss},\omega_{\rm ss})$.

We are interested in solutions with $\Gamma$ close to $1$. 
In what follows it is convenient to define a new parameter
\be
  k = \sqrt{\Gamma-1}\ .
\n{eq3.19}
\ee

Our strategy is to expand the CSS solutions in powers of $k$. 
We can deduce the leading terms by looking at solutions for $k$ close to zero.
Our ansatz is
\be
N_{\rm ss} = \frac{\bar{N}_0 e^{-x}}{k}\ ,
\n{eq3.20}
\ee
\be
A_{\rm ss}(x) = 1+\bar{A}(x)\, k^2\ ,
\n{eq3.21}
\ee
\be
w_{\rm ss}(x) = \bar{\omega}(x)\, k^2\ ,
\n{eq3.22}
\ee
\be
v_{\rm ss}(x) = \bar{v}(x)\, k\ .
\n{eq3.23}
\ee
Substituting the expressions \eref{eq3.19}--\eref{eq3.23}
into the equations (\ref{eq3.13})--(\ref{eq3.18})
and keeping only the leading terms we obtain the equations for $\bar{\omega}$ and $\bar{v}$
\be
  (1+\bar{N}_0 e^{-x}\bar{v})\bar{\omega}_{,x} + \bar{N}_0 e^{-x}\bar{\omega}\bar{v}_{,x} = 0\ ,
\n{eq3.24}
\ee
\be
  \frac{\bar{N}_0}{\bar{\omega}}\bar{\omega}_{,x} + (e^x+\bar{N}_0\bar{v})\bar{v}_{,x} = 
  -\bar{N}_0(\bar{\omega}+\bar{N}_0\bar{\omega}\bar{v}e^{-x}-2)\ .
\n{eq3.25}
\ee
The algebraic equation \eref{eq3.18} yields an expression for $\bar{A}$
\be
  \bar{A} = 2\bar{\omega}(1 + \bar{N}_0e^{-x}\bar{v})\ ,
\n{eq3.26}
\ee
which we used to eliminate $\bar{A}$ from the equations.

The scaling relations (\ref{eq3.20})--(\ref{eq3.23}) are those of Newtonian
CSS solutions. Indeed, the equations (\ref{eq3.24})--(\ref{eq3.25}) are exactly
the CSS Newtonian equations as can be seen by comparison with
equations $(3.5)$--$(3.7)$ of \cite{Ori_Piran_90} after the transformation 
$\bar{x} = -e^x/\bar{N}_0$.

\begin{figure}
\begin{center}
\epsfig{file=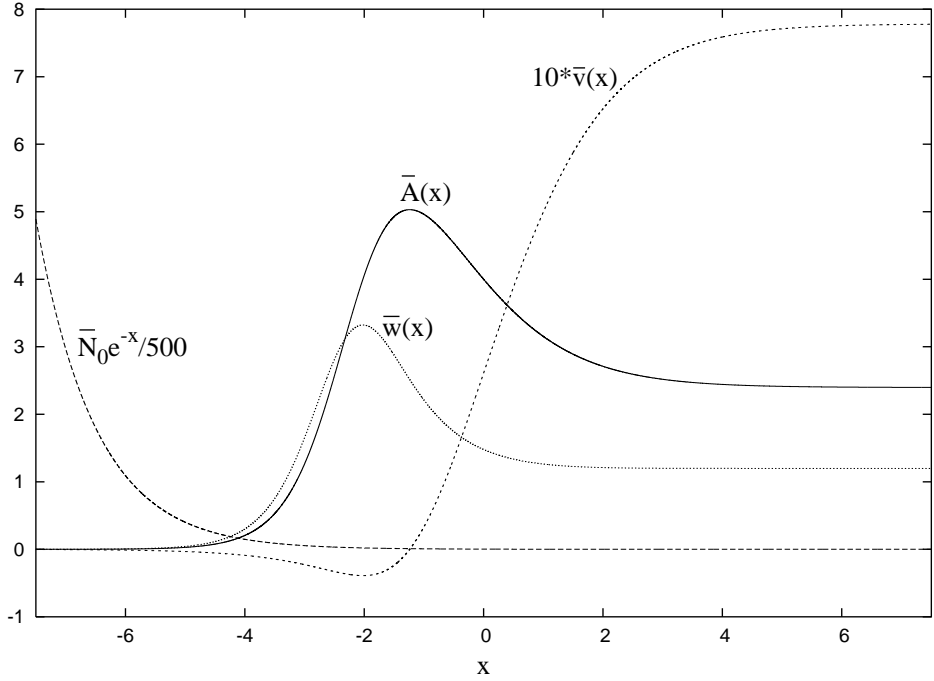, width=\textwidth}
\end{center}
\caption{Plot of the limiting solutions of $\bar{N}=\bar{N}_0 e^{-x}$, $\bar{A}(x)$, 
$\bar{\omega}(x)$, $\bar{v}(x)$.}
\label{fig:css}
\end{figure}

The numerical solutions of the equations (\ref{eq3.24})--(\ref{eq3.25}) is
shown in figure~\ref{fig:css}.
In this solution the velocity crosses the zero line once and it is the critical solution.
It is the so-called Hunter's solution of type (a) \cite{Hunter_77}.

\section{Perturbations of the CSS solutions}
\label{sec:perturb}
\setcounter{equation}0

Again, following \cite{Hara_Koike_Adachi96} we calculate the relevant perturbations using 
a linear expansion
\be
  H(x,s) = H_{\rm ss}(x) + \epsilon\,h_{\rm var}(x,s)\ ,
\n{eq4.1}
\ee
where $H(x,s)$ is one of $\{\log(N),\log(A),\log(w),v\}$ and
$h_{\rm var}(x,s)$ has the form
\be
  h_{\rm var}(x,s) = h_{\rm p}(x) e^{\kappa s} \hspace{2cm}\kappa\in \mathbb{C}\ .
\n{eq4.2}
\ee

The equations for $h_{\rm p}(x)$ are rather lengthy and can be found in \cite{Hara_Koike_Adachi96}
so we do not reproduce them here.
Using the same strategy as in the previous section, we expand $h_{\rm p}(x)$ and
$\kappa$ in powers of $k$. 
The observed behavior is 
\be
N_{\rm p}(x) = \bar{N}_{\rm p}(x)\ ,
\n{eq4.3}
\ee
\be
A_{\rm p}(x) = \bar{A}_{\rm p}(x)\ ,
\n{eq4.4}
\ee
\be
w_{\rm p}(x) = \frac{\bar{w}_{\rm p}(x)}{k^2}\ ,
\n{eq4.5}
\ee
\be
v_{\rm p}(x) = \frac{\bar{v}_{\rm p}(x)}{k}\ ,
\n{eq4.6}
\ee
\be
\kappa = \bar{\kappa} + O(k^2)\ .
\n{eq4.7}
\ee
After substituting (\ref{eq3.20})--(\ref{eq3.23}) and  (\ref{eq4.3})--(\ref{eq4.7})
 into the original equations
and keeping only the leading terms we obtain a set of ODE's
\[
\fl
  \left[
  \left(\begin{array}{cccc}
           1 & 0 & 0 & 0\\
           0 & 1 & 0 & 0\\
           0 & 0 & 1+\bar{N}_0 e^{-x}\bar{v} & \bar{N}_0 e^{-x}\\
           0 & 0 & \bar{N}_0 e^{-x} & 1+\bar{N}_0 e^{-x}\bar{v}
         \end{array}\right)\derv{}{x}\right.
\]
\[
         -
  \left.
  \left(\begin{array}{cccc}
           -1 & 0 & 2\bar{\omega} & 0\\
           1 & 0 & -\bar{\omega} & 0\\
           0 & 0 & 0 & -\frac{\bar{N}_0 e^{-x}\bar{\omega}_{,x}}{\bar{\omega}}\\
           -\frac{\bar{N}_0 e^{-x}}{2} & 0 & 0 &-\bar{N}_0 e^{-x}\bar{v}_{,x}
         \end{array}\right)
  \right]
  \left(\begin{array}{c}
    \bar{A}_{\rm p}\\
    \bar{N}_{\rm p}\\
    \bar{w}_{\rm p}\\
    \bar{v}_{\rm p}
  \end{array}\right)
  =
\]
\be
  -\bar{\kappa}
  \left(\begin{array}{cccc}
           0 & 0 & 0 & 0\\
           0 & 0 & 0 & 0\\
           0 & 0 & 1 & 0\\
           0 & 0 & 0 & 1
         \end{array}\right)
  \left(\begin{array}{c}
    \bar{A}_{\rm p}\\
    \bar{N}_{\rm p}\\
    \bar{w}_{\rm p}\\
    \bar{v}_{\rm p}
  \end{array}\right)\ .
\n{eq4.8}
\ee
The algebraic equation has the form
\be
  (1-\bar{\kappa})\bar{A}_{\rm p}
  -2\bar{N}_0 e^{-x}\bar{\omega}(\bar{v}_{\rm p}+
  \bar{v}\,\bar{\omega}_{\rm p})
  -2\bar{\omega}\bar{\omega}_{\rm p} = 0\ ,
\n{eq4.9}
\ee
which can be used to verify the consistency of the results.

\begin{figure}[h]
\begin{center}
\epsfig{file=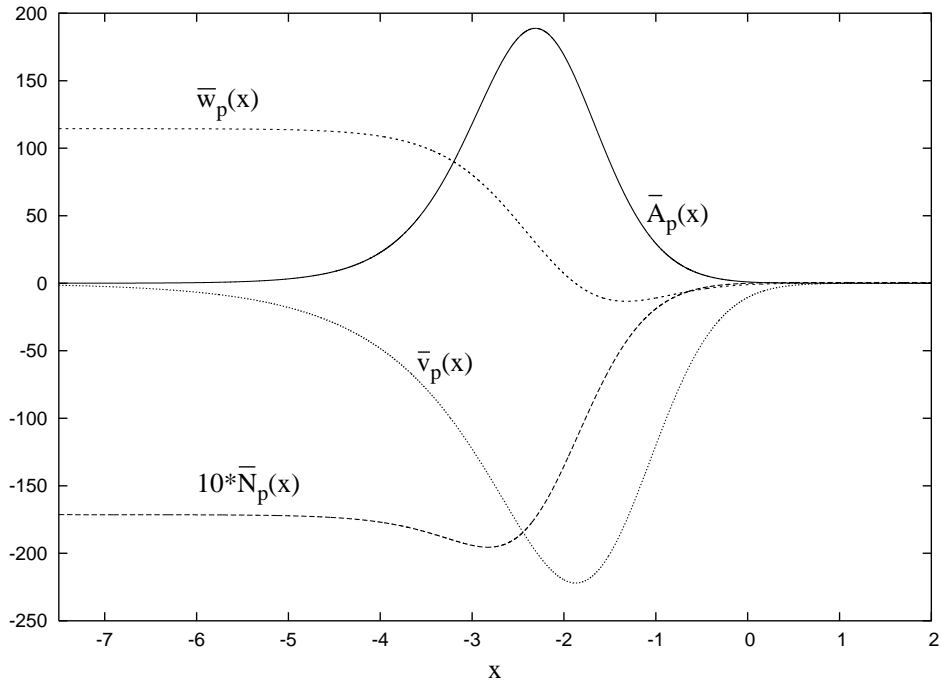, width=\textwidth}
\end{center}
\caption{Plot of the relevant eigenmodes of the limiting CSS solution
$\bar{N}_{\rm p}(x)$, $\bar{A}_{\rm p}(x)$,
$\bar{\omega}_{\rm p}(x)$, $\bar{v}_{\rm p}(x)$.}
\label{fig:css_pert}
\end{figure}

The solution methods for the limiting case are the same as described
in \cite{Hara_Koike_Adachi96}.
In particular to solve for the $\bar{v}(x)$ and $\bar{\omega}(x)$,
we must regularize the equations (\ref{eq3.24})--(\ref{eq3.25}) at the sonic point ($x=0$)
and find the value of $\bar{v}(0)$ for which the solution has the correct behavior
as $x\to -\infty$.
Similarly, the value of $\bar{\kappa}$ is found by requiring that 
$\bar{v}_{\rm p}(x)$ does not diverge as $x\to -\infty$.
To numerically integrate the set of ODE's we used the package LSODE \cite{lsode1,lsode2}.
For increased precision we used $128$ bit representation of floating point numbers
(quadruple precision).
Figure \ref{fig:css_pert} shows the limiting solutions for the relevant eigenmodes.

\section{Results of numerical calculations}
\label{sec:results}
Numerical modeling of type II critical collapse of relativistic 
fluids even in spherical symmetry is a rather challenging task.
The reason is that the dynamics takes place at ever decreasing spatial 
length scale and also the density and pressure of the fluid increase 
by many orders of magnitude.
For example, in our calculations the density and pressure increase typically 
by a factor of million for the nearly critical solutions (in the subcritical regime)
and in some of the supercritical calculations the central density reached a value
 greater than $10^{54}$.
It is clear that some kind of adaptivity is required in order to perform 
these calculations.

For $\Gamma$ close to $1$, i.e., in the regime we want to explore, another
difficulty arises --- tuning the critical parameter to about 15 digits
(the limit for double precision floating point numbers) is not sufficient
and calculations in quadruple precision are necessary.
This prolongs the runtime of the code significantly because on ``standard''
workstations based on the x86 architecture quadruple precision is not
natively supported by the processor.
A more detailed description of some of the specifics of our numerical implementation
can be found in section~\ref{sec:numerics}. 

\subsection{Comparison of numerical and CSS critical solutions}

If our numerical scheme works properly we should obtain the same critical 
solutions as we did by using the CSS ansatz (up to a truncation error).
We must bear in mind, however, that the CSS critical solutions do not
describe an asymptotically flat spacetime whereas the spacetime generated
numerically is asymptotically Schwarzschild and therefore the solutions
match only in a limited domain close to $r=0$.
In order to compare the solutions we have to translate the coordinate $r$
used in numerical calculations into the self similar coordinate $x$ in which
the CSS solutions are cast.
The relation is provided by the equation \eref{eq3.3}.
Since the relation of $t_*$ to $t$ is not known we use some distinct feature
of the solution, e.g., local minimum or maximum to identify a particular
$r$ with a particular $x$.
This allows us to calculate $t_*$.

\begin{figure}
\begin{center}
\epsfig{file=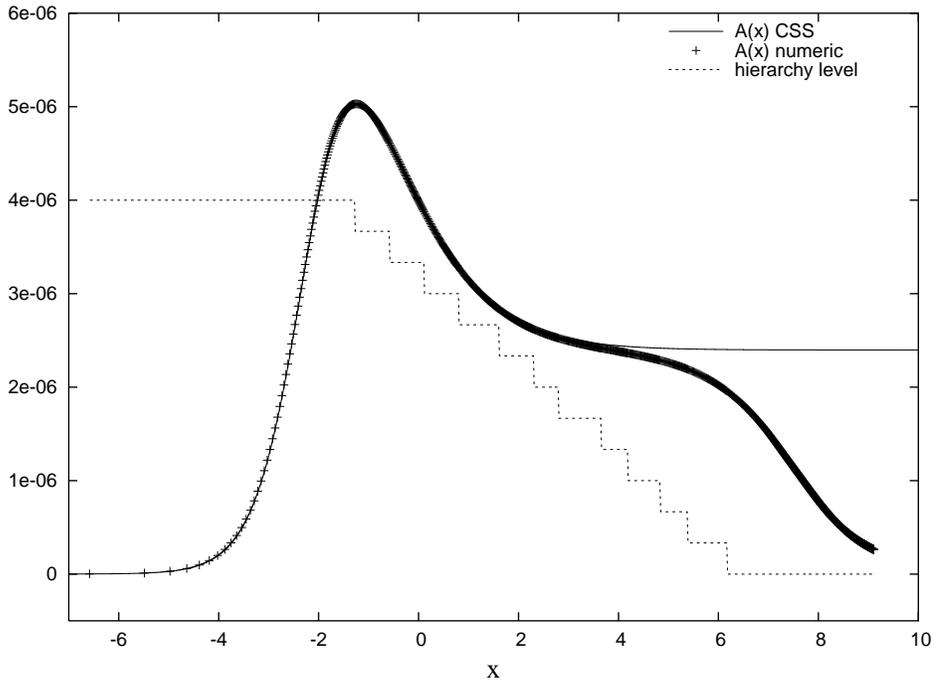, width=\textwidth}
\end{center}
\caption{Comparison of CSS and numerical solution for $A(x)$ for 
$k^2=10^{-6}$.}
\label{fig:A_comp}
\end{figure}

\begin{figure}
\begin{center}
\epsfig{file=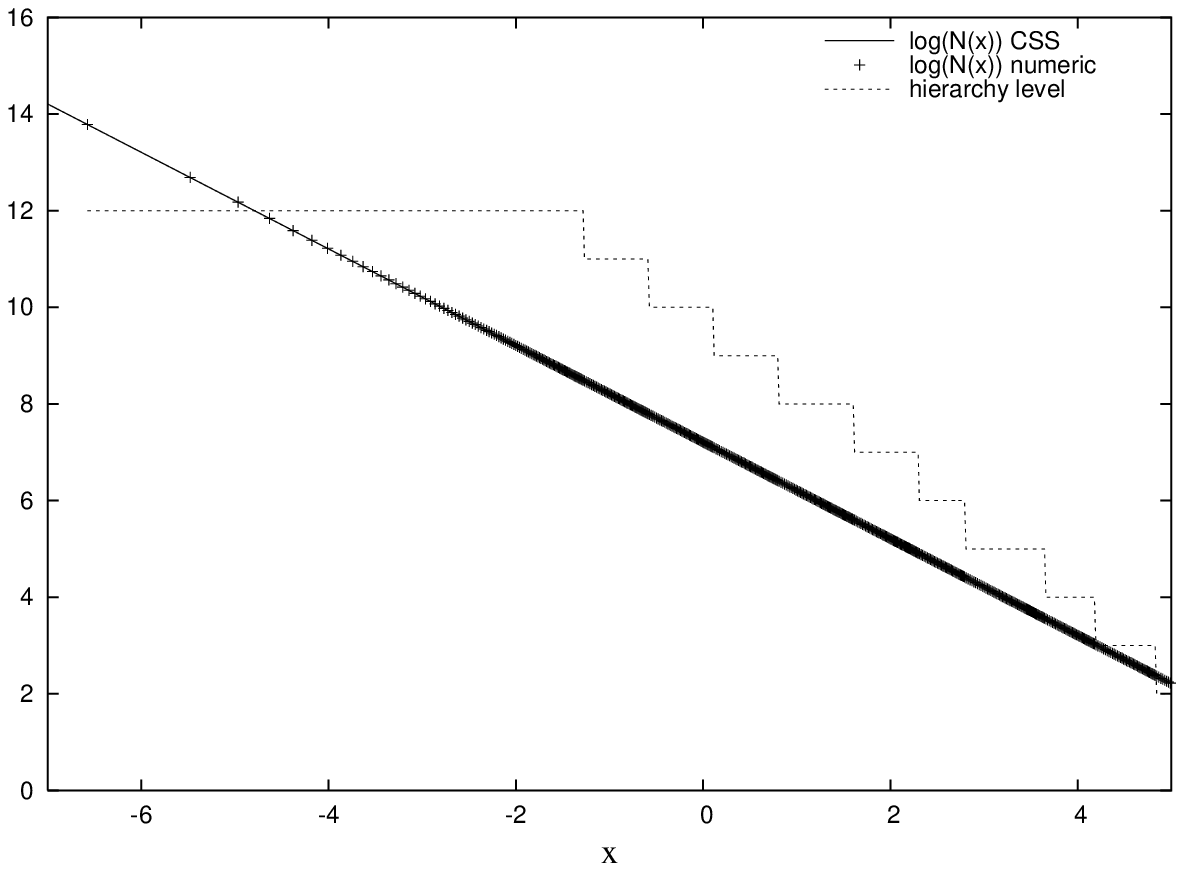, width=\textwidth}
\end{center}
\caption{Comparison of CSS and numerical solution for $\log(N(x))$ for
$k^2=10^{-6}$.}
\label{fig:logN_comp}
\end{figure}

\begin{figure}
\begin{center}
\epsfig{file=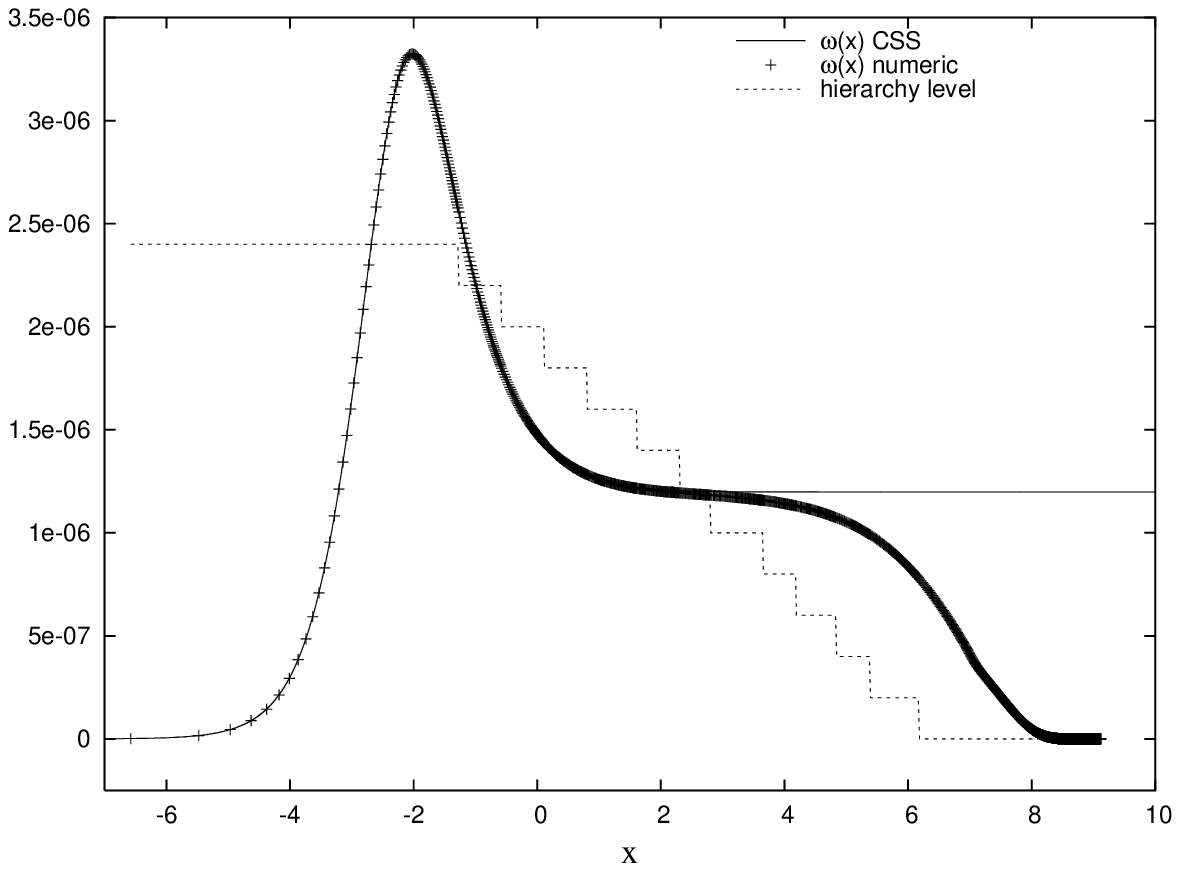, width=\textwidth}
\end{center}
\caption{Comparison of CSS and numerical solution for $\omega(x)$ for
$k^2=10^{-6}$.}
\label{fig:w_comp}
\end{figure}

\begin{figure}
\begin{center}
\epsfig{file=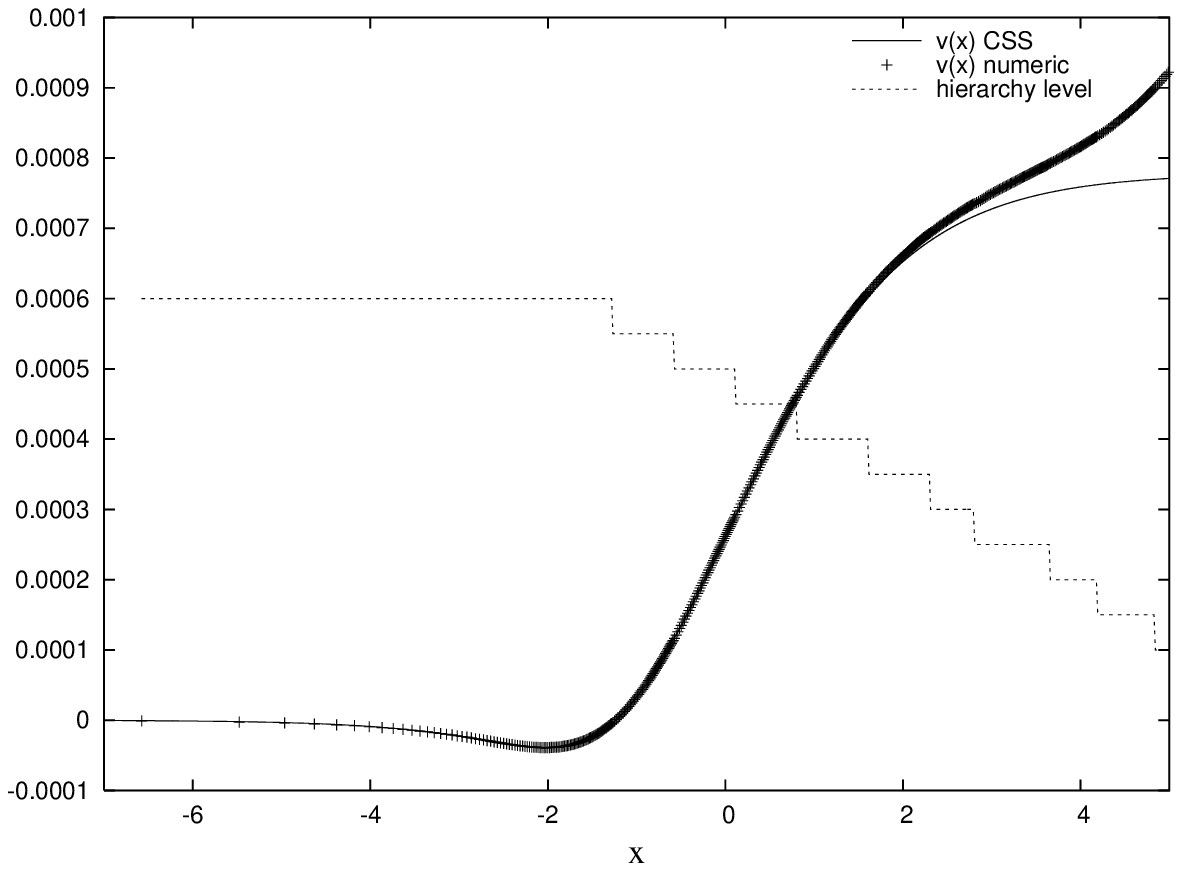, width=\textwidth}
\end{center}
\caption{Comparison of CSS and numerical solution for $v(x)$ for
$k^2=10^{-6}$.}
\label{fig:v_comp}
\end{figure}

Figures \ref{fig:A_comp}--\ref{fig:v_comp} show the comparison for 
$k^2=10^{-6}$.
The numerical data were taken from the closest subcritical run just 
before the fluid dispersed.
The numerical solutions agree very well with the CSS ones in a 
limited region as expected.
The results for other values of $k$ agree similarly well.

Note that the lapse $\alpha$ used in the numerical calculations is not 
the same as the $\alpha$ from equation \eref{eq3.4}.
Therefore in order to compare the functions $N(x)$ we must rescale 
one by a constant factor --- we simply matched the leftmost data point from 
the numerical calculations with the corresponding one from the CSS solution.
The dotted line shows the AMR hierarchy level. An increase in hierarchy level
by one corresponds to a reduction of the cell size by factor of $2$.

\subsection{Determining the scaling exponent}

For the numerical calculations we chose the initial density profile to be a 
Gaussian 
\be
  \rho = p \exp{\left[-(r-r_c)^2/\Delta^2\right]}\ ,
\n{eq5.10}
\ee
and the velocity $v$ was set to zero (we used $r_c=0$ and $\Delta=0.2$).
The amplitude of the Gaussian $p$ serves as the tunable parameter.
Its critical value $p_*$ can be found by a bisection search.
Once the critical parameter is obtained with sufficient precision 
it is possible to calculate the scaling exponent.

We calculate the scaling exponent from {\em subcritical} runs.
It has the obvious advantage that a black hole does not need to form
in our spacetime (in our coordinate system it is not even possible).
A further explanation will be given in the next section.
The scaling relation for the trace of the stress-energy tensor 
has the form
\be
  \max(T^{\mu}_{\ \ \mu}) = 3P-\rho\sim |p-p_*|^{-2\gamma}\ .
\n{eq5.11}
\ee

An accurate determination of the scaling exponent in practice 
is not as straightforward as it might seem. 
The basic idea is to perform a number of subcritical runs and then fit
a straight line to the logarithm of the equation \eref{eq5.11}.
In our approach we do not use a predetermined value of $p_*$ but also 
optimize $p_*$ in order to get the best fit. 
In that sense the fit is non-linear.

We would like to have our data points separated approximately 
evenly in the $\log|p-p_*|$
coordinate but since we do not know the value of $p_*$ beforehand 
we perform a trial fit to obtain the value of $p_*$ and then use 
that value as a reference.

Another problem is to determine the interval of $\log|p-p_*|$ over which we 
perform the fit. 
A very broad interval is likely to give inaccurate results since the scaling
behavior is valid only in the vicinity of $p_*$.
On the other hand fitting over a very narrow interval extremely close to the $p_*$
might produce incorrect result as well.
We use the following approach which we believe is sufficiently robust and 
provides a way to estimate the error as well.

First we generate data points that span rather wide range of $\log|p-p_*|$.
Typically, we use around one hundred data points.
Then we perform what we call a {\em windowed fit}, i.e., fit data points 
with the index $i=k,\dots,k+N_{\rm w}-1$ where $k=1,\dots,N-N_{\rm w}+1$.
$N_{\rm w}$ is the width of the window and $N$ is the total number of data points.
If $N_{\rm w}=N$ then the window spans all the data points.

\begin{figure}
\begin{center}
\epsfig{file=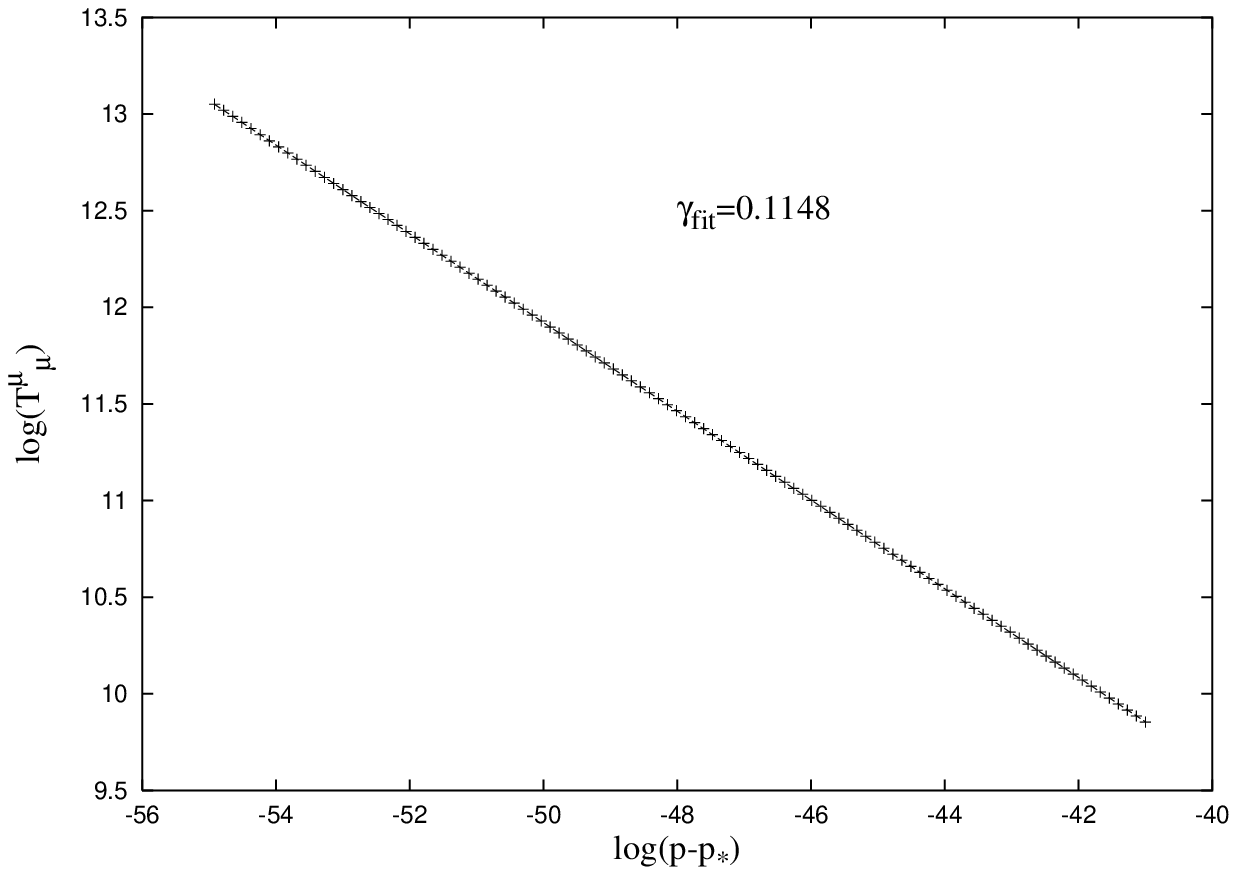, width=0.95\textwidth}
\end{center}
\caption{Fitted data from  subcritical solutions for $k^2=10^{-2}$.}
\label{fig:total_01}
\end{figure}

\begin{figure}
\begin{center}
\epsfig{file=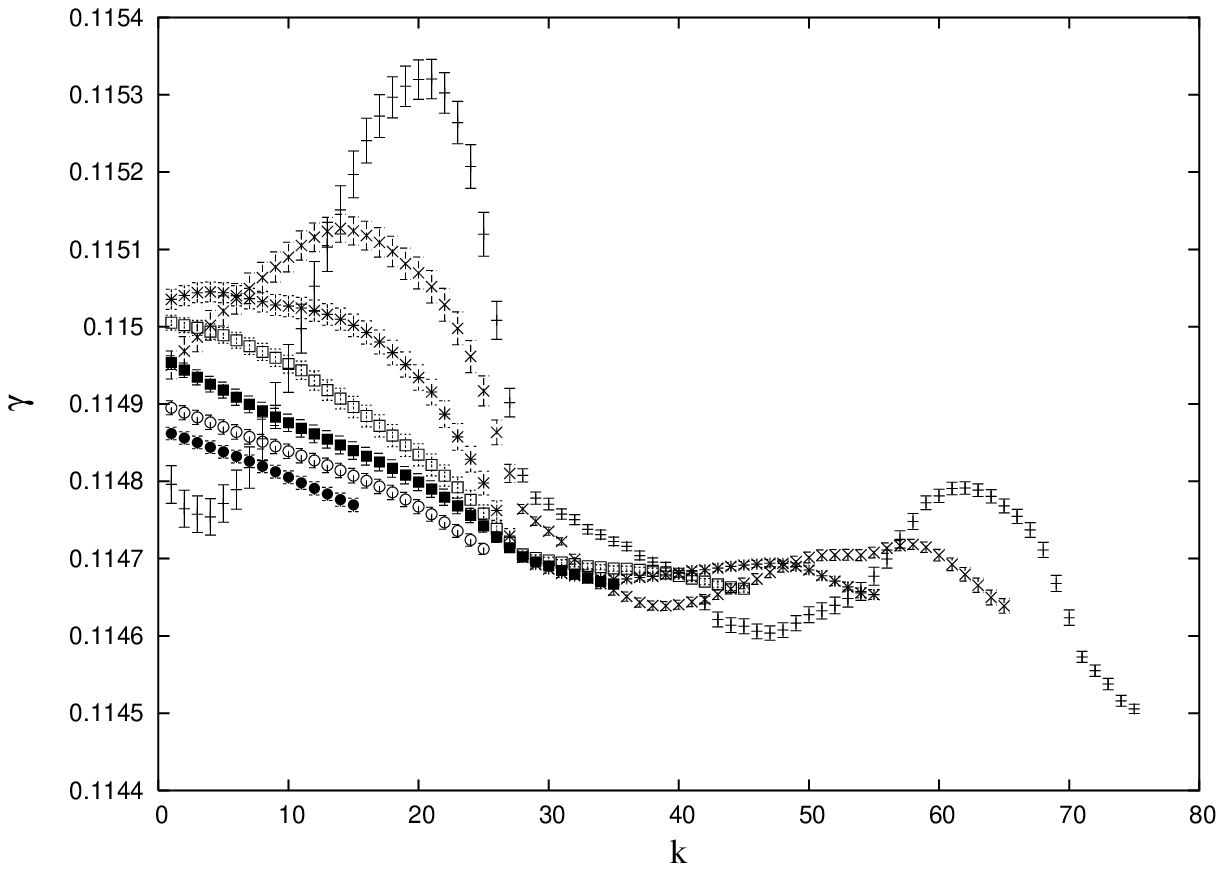, width=0.95\textwidth}
\end{center}
\caption{Windowed fits of data from subcritical runs for $k^2=10^{-2}$,
         $N_{\rm w}=30,40,50,60,70,80,90$.}
\label{fig:windowed_01}
\end{figure}

\begin{figure}
\begin{center}
\epsfig{file=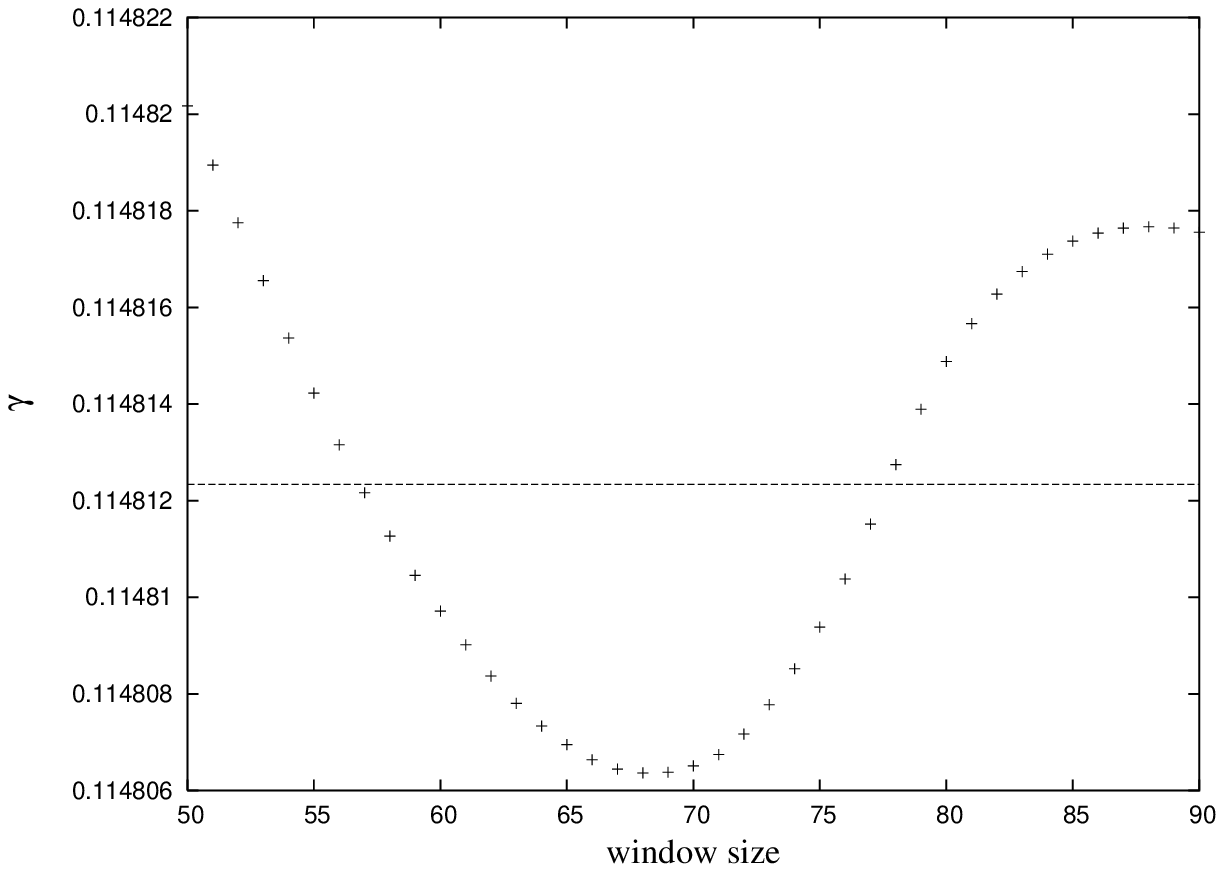, width=0.95\textwidth}
\end{center}
\caption{Averaged $\gamma$ as a function of window size for $k^2=10^{-2}$.}
\label{fig:windowed_gamma01}
\end{figure}

Figure \ref{fig:total_01} shows data obtained from subcritical 
runs for $k^2 = 10^{-2}$.
The dashed line (not very visible) corresponds to the best fit
and yields a value of $\gamma_{\rm fit} = 0.1148$.
This is the value we report.
Figure \ref{fig:windowed_01} shows $\gamma$ as a function of
$k$ for various window sizes (different markers correspond to 
different window sizes).
Each data point is plotted with an error bar corresponding to the 
error of each $\gamma$ estimate that is equal to one half of the
standard deviation corresponding to the slope $-2\gamma$.
The smaller the error bar the better the linear fit.
The number of data points for each window size corresponds to the
largest $k$ of $N-N_{\rm w}+1$.
The main purpose of this plot is to provide some quantitative 
estimate of the error of the $\gamma_{\rm fit}$.
In addition, it may provide a hint whether there is some trend in $\gamma$
as we fit data closer to the critical value $p_*$.
Obviously, the largest variations are present for the smallest window sizes
since we fit less data points.
However, the averages
\be
 \gamma_{\rm av} = \frac{1}{N-N_{\rm w}+1}\sum^{N-N_{\rm w}+1}_{k=1}\gamma_{\rm k} 
\n{eq5.12}
\ee
for different window sizes $N_{\rm w}$ show very little variation.
This is illustrated in figure \ref{fig:windowed_gamma01}.

Looking at figure \ref{fig:windowed_01} (although rather messy) we can
very conservatively declare that for $k^2=10^{-2}$
\be
  \gamma < 0.1153\ . 
\n{eq5.13}
\ee
The plot also suggest a trend of decreasing $\gamma$ as 
we fit data closer to $p_*$. 
Therefore it is not straightforward to estimate the lower bound for $\gamma$.

\begin{figure}
\begin{center}
\epsfig{file=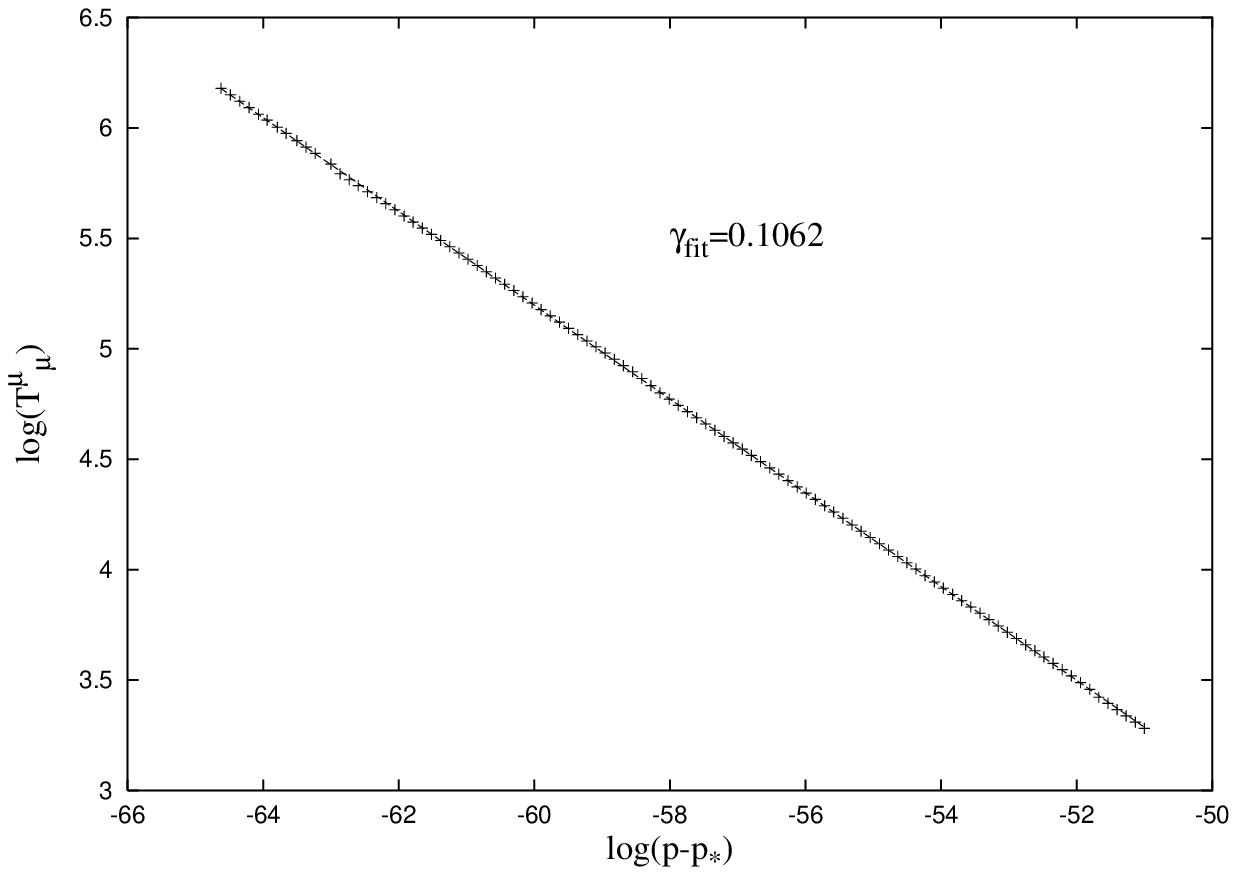, width=0.95\textwidth}
\end{center}
\caption{Fitted data from  subcritical solutions for $k^2=10^
{-5}$.}
\label{fig:total_00001}
\end{figure}

\begin{figure}
\begin{center}
\epsfig{file=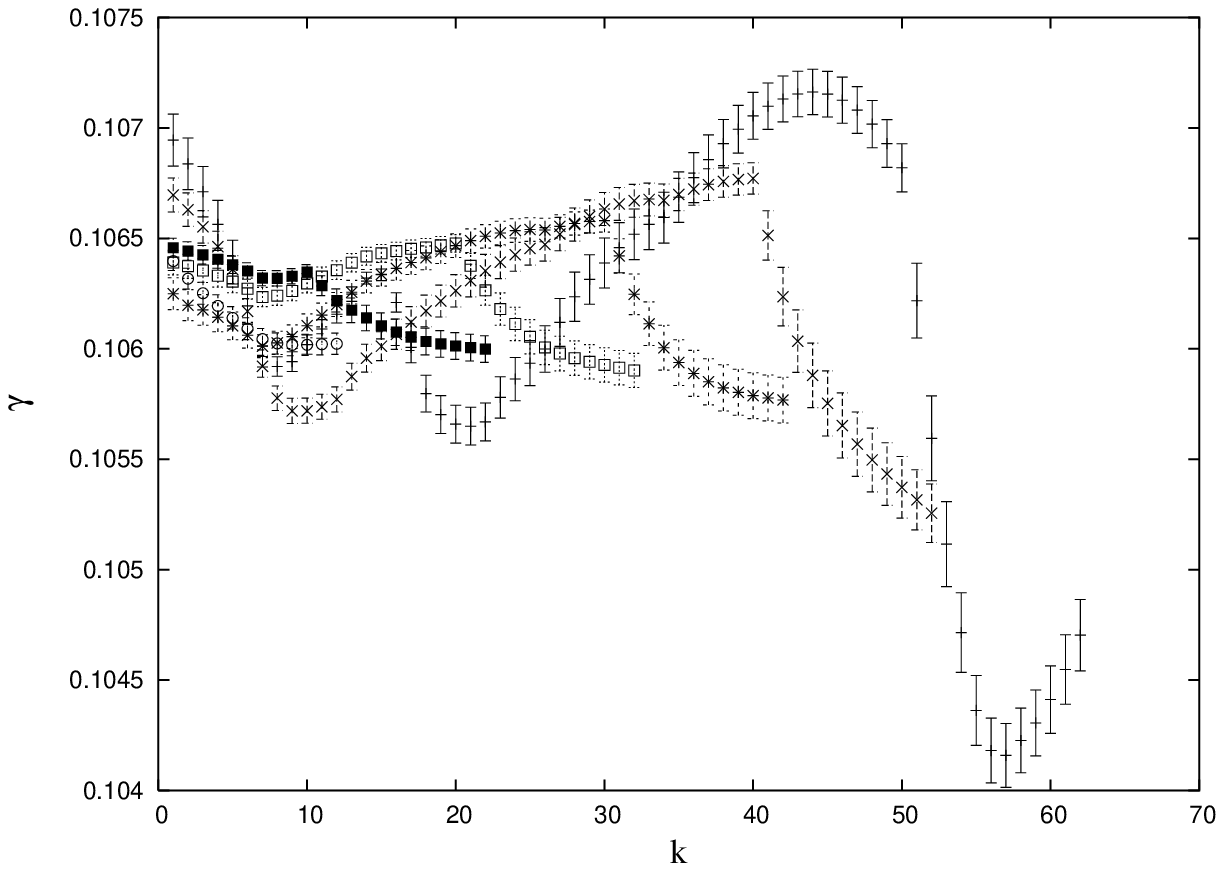, width=0.95\textwidth}
\end{center}
\caption{Windowed fits of data from subcritical runs for $k^2
=10^{-5}$,
         $N_{\rm w}=40,50,60,70,80,90$.}
\label{fig:windowed_00001}
\end{figure}

\begin{figure}
\begin{center}
\epsfig{file=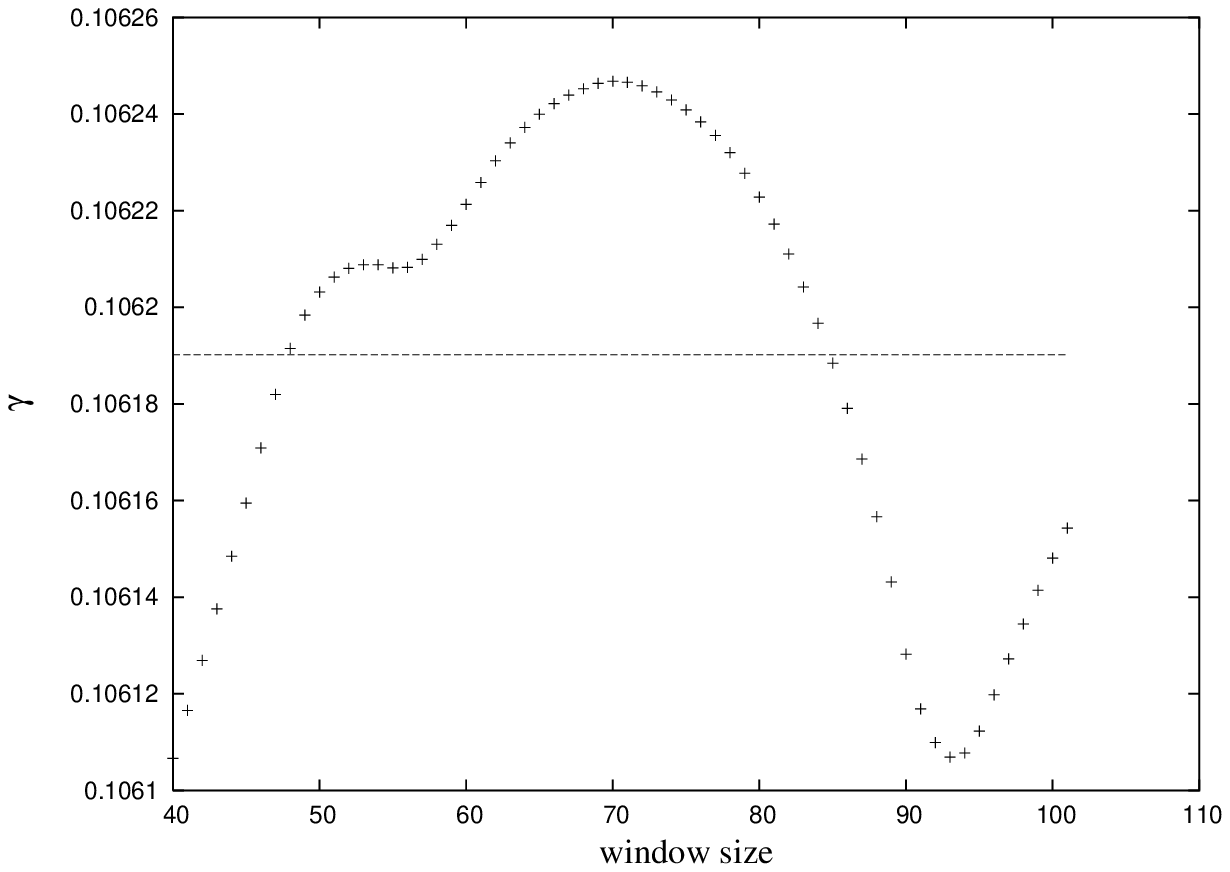, width=0.95\textwidth}
\end{center}
\caption{Averaged $\gamma$ as a function of window size for $
k^2=10^{-5}$.}
\label{fig:windowed_gamma00001}
\end{figure}

We created the same type of plots for all the other values of $k$. 
Their structure is similar therefore we only show one additional 
set for $k^2=10^{-5}$ (figures \ref{fig:total_00001}--\ref{fig:windowed_gamma00001}).
For $k^2=10^{-5}$ we do not see any obvious decreasing trend  for $\gamma$.
Moreover, we observe a drop in the $\gamma$ value for the smallest window
 sizes plotted and large values of $k$.
This just illustrates the fact that it could be dangerous to rely only on data
from a narrow interval around the critical value $p_*$.

Table~\ref{table:results} summarizes the results obtained from both
 the perturbation theory ($\gamma_{\rm ss}$) and the numerical calculations
($\gamma_{\rm fit}$).
The error reported is the percentage difference between $\gamma_{\rm ss}$ and
$\gamma_{\rm fit}$.
Our results are in perfect agreement with \cite{Hara_Koike_Adachi96} for the
values of $k$ reported therein. 

\begin{table}
\begin{center}
\begin{tabular}{|c|c|c|c|c|}\hline
$k^2$ &$\kappa$ & $\gamma_{\rm ss}$ & $\gamma_{\rm fit}$&{\rm error}(\%) \\ \hline\hline
$10^{-2}$&$8.748687152$&$0.1143028643$&$0.1148$&0.4\\
$10^{-3}$&$9.386603219$&$0.1065348110$&$0.1071$&0.5\\
$10^{-4}$&$9.455924881$&$0.1057538012$&$0.1062$&0.4\\
$10^{-5}$&$9.462917038$&$0.1056756596$&$0.1062$&0.5\\
$10^{-6}$&$9.463616859$&$0.1056678451$&$0.1064$&0.7\\
      $0$&$9.463694624$&$0.1056669768$&        &   \\ \hline
\end{tabular}
\caption{Scaling exponents calculated using both the CSS ansatz and a
direct numerical solution. The values reported for $k=0$ are the limiting values.}
\label{table:results}
\end{center}
\end{table}

\subsection{The supercritical regime}

In the previous section we showed how to calculate the critical exponents
from subcritical runs using \eref{eq5.11}.
Originally, the order parameter was taken to be the mass of a black hole formed 
during the supercritical collapse.
Why haven't we then used the supercritical runs and the black hole mass to 
determine the scaling exponent?
Aside of the fact that in the coordinate system we are using we can not really 
form a black hole a more fundamental reason exists.
There is no sign of black hole formation in the supercritical runs.
Even after the central densities reach large values, there is no sign
of event horizon formation (i.e. $2m/r$ approaches a constant value smaller than $1$). 
According to \cite{Harada_Maeda_2001} this is expected and the universal
attractor is not a spacetime with a black hole but a general relativistic Larson-Penston
solution (GRLP).
The GRLP solution exists only for $\Gamma-1<0.036\pm 0.002$ 
and it contains a naked singularity for $\Gamma-1<0.0105$ \cite{Ori_Piran_90}.

The GRLP is a solution of equations \eref{eq3.13}--\eref{eq3.17} but unlike
the critical solution (which is a general relativistic generalization of the 
Hunter's type (a) solution) it is a ``pure collapse'' solution, i.e., the velocity
is always negative.

To test this hypothesis, we performed a generic supercritical run for $\Gamma-1=0.01$
and $\Gamma-1=10^{-6}$.
For $\Gamma-1=10^{-6}$ we stopped the calculations when the refinement level reached $100$.
At that point the central density reached value of about $10^{54}$ 
and the spatial resolution at the center was about $10^{-32}$.
For $\Gamma-1=0.01$ we stopped the calculations at refinement level $65$.
The central density reached value of about $10^{38}$
and the spatial resolution at the center was about $10^{-22}$.
We stress that the calculations were stopped artificially and they could be, if needed, 
pushed further.
We also performed a control supercritical run for $\Gamma-1 = 0.02$ and we observed
the quantity $2m/r$ approaching  $1$ as expected.

\begin{figure}[t]
\begin{center}
\epsfig{file=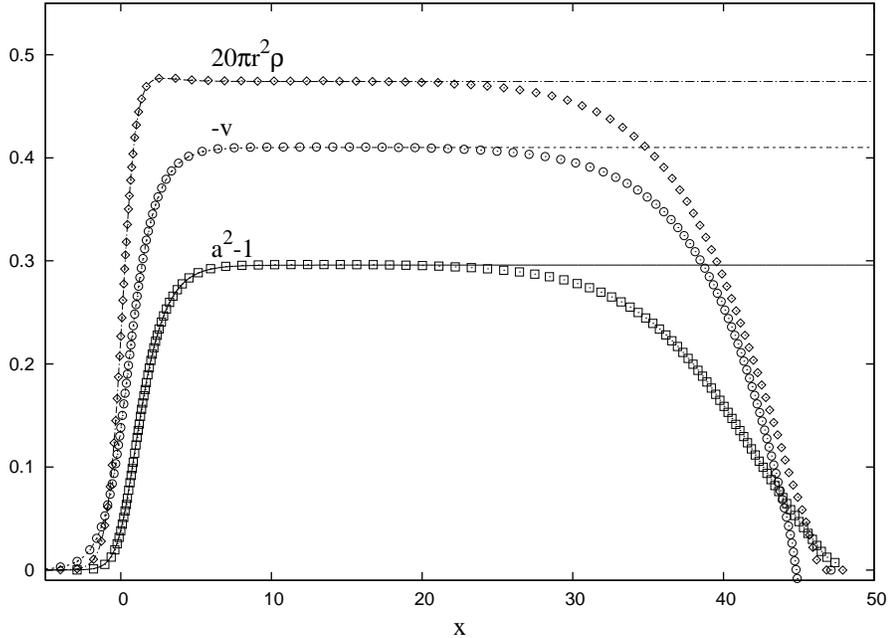, width=0.95\textwidth}
\end{center}
\caption{Comparison of the supercritical numerical solution and
 the GRLP solution for $\Gamma-1 = 0.01$.}
\label{fig:super_whole_01}
\end{figure}

\begin{figure}[h]
\begin{center}
\epsfig{file=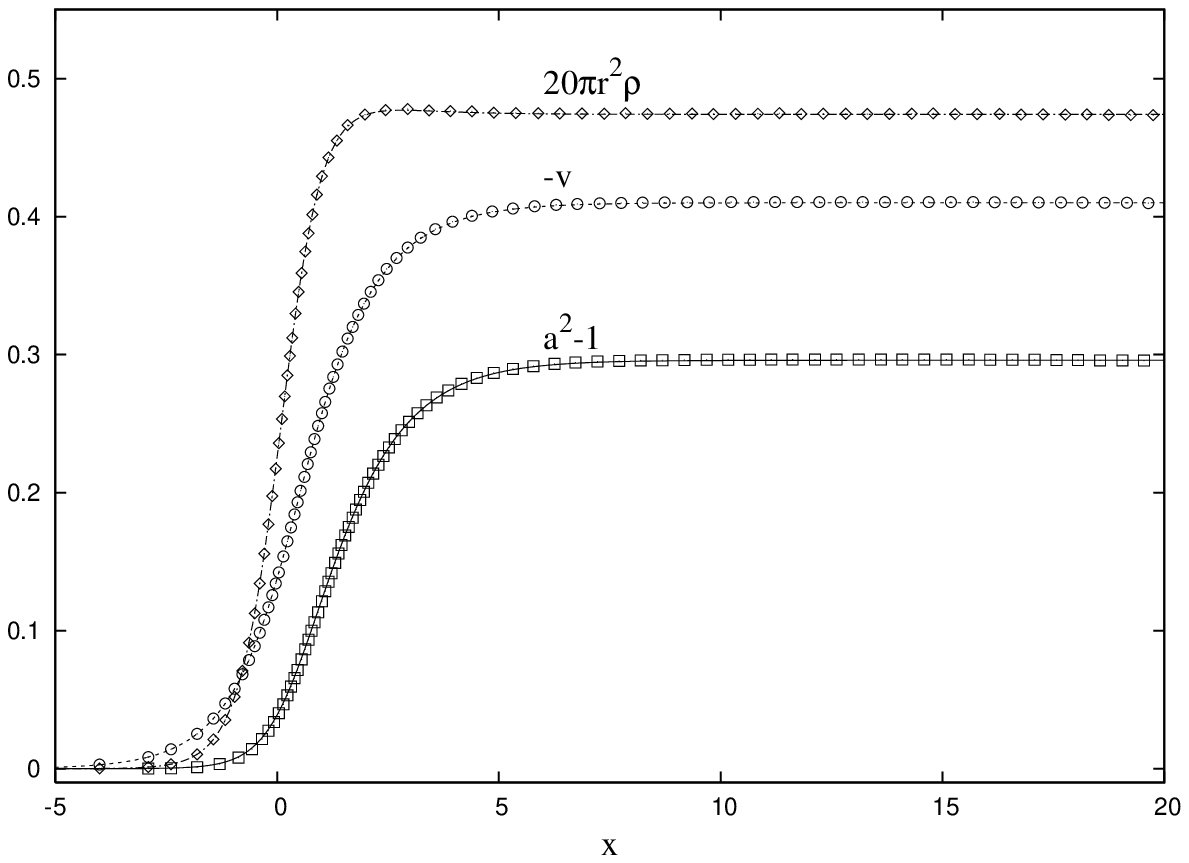, width=0.95\textwidth}
\end{center}
\caption{Comparison of the supercritical numerical solution and
 the GRLP solution for $\Gamma-1 = 0.01$ --- detail of the central region.}
\label{fig:super_central_01}
\end{figure}

\begin{figure}[t]
\begin{center}
\epsfig{file=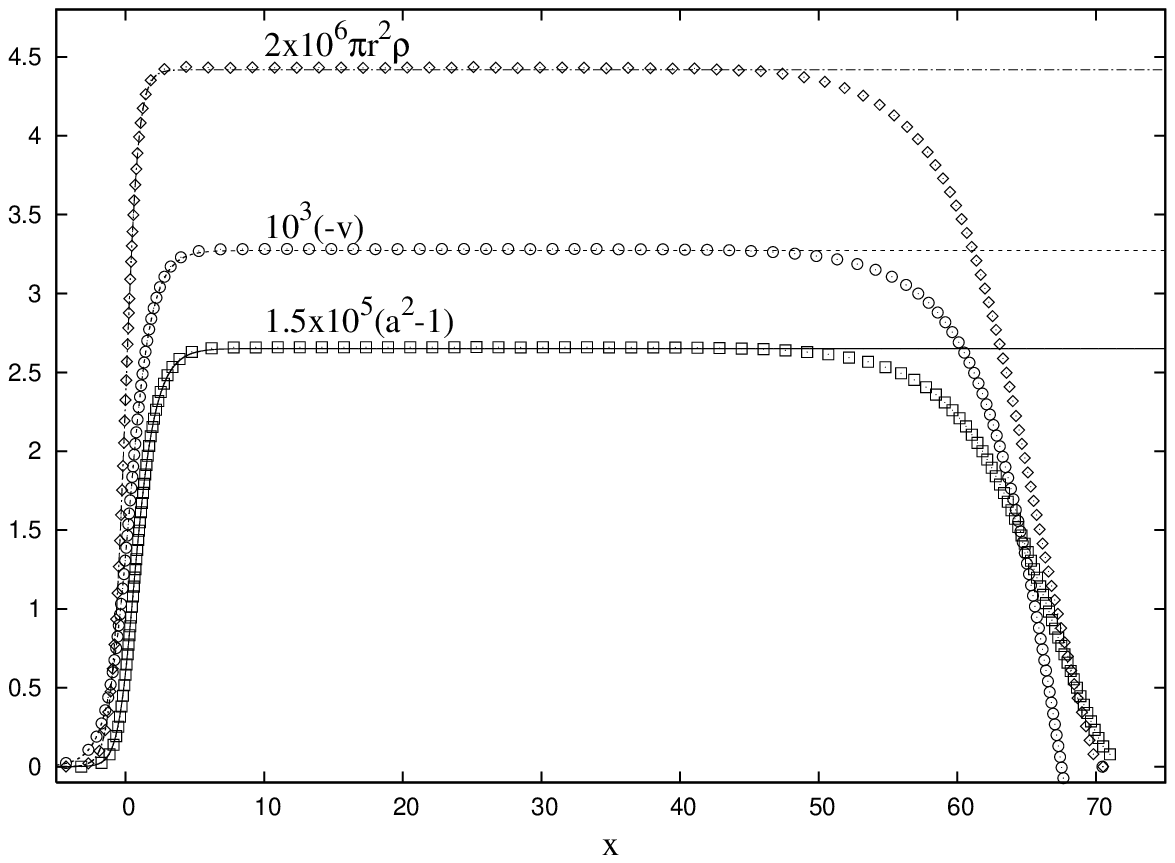, width=0.95\textwidth}
\end{center}
\caption{Comparison of the supercritical numerical solution and
 the GRLP solution for $\Gamma-1 = 10^{-6}$.}
\label{fig:super_whole_000001}
\end{figure}

\begin{figure}[h]
\begin{center}
\epsfig{file=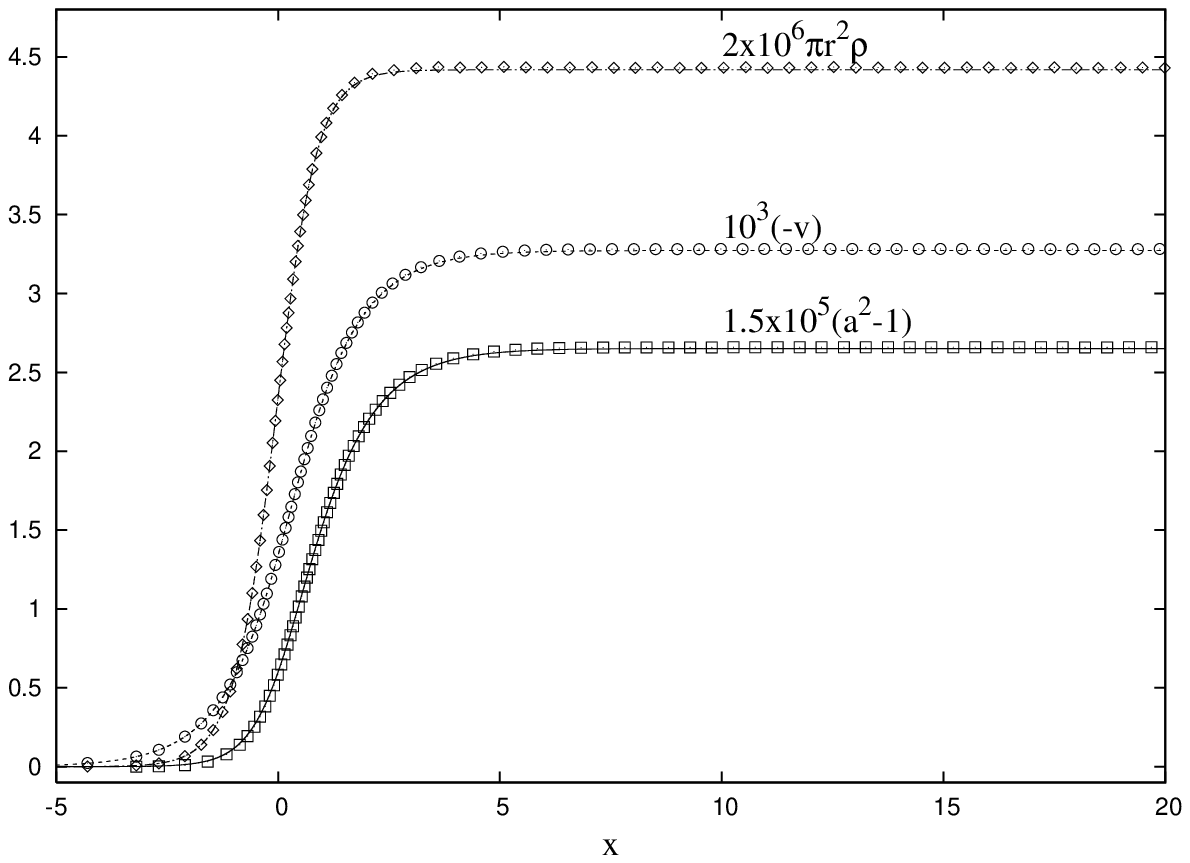, width=0.95\textwidth}
\end{center}
\caption{Comparison of the supercritical numerical solution and
 the GRLP solution for $\Gamma-1 = 10^{-6}$ --- detail of the central region.}
\label{fig:super_central_000001}
\end{figure}

Figures \ref{fig:super_whole_01} and \ref{fig:super_central_01} show the comparison of the GRLP solution with the numerical data for $\Gamma-1=0.01$. 
Figures \ref{fig:super_whole_000001} and \ref{fig:super_central_000001} 
show the comparison of the GRLP solution with the numerical data for $\Gamma-1=10^{-6}$.
In the plots we show only a fraction of the data points
outside of the very central region so the GRLP solution is visible on the plots.
The agreement is very good near the origin and starts to deviate at larger distances.
This is expected since the GRLP solution is not asymptotically flat whereas
the numerical solution is.

These results support the hypothesis that the GRLP solution is an attractor for the
supercritical collapse and therefore naked singularities can be formed in a 
generic gravitational collapse.
It is conceivable that black hole is also an attractor for the supercritical collapse.
It would certainly be interesting to investigate the 
regime of the transition from GRLP to a black hole spacetime.

\section{Numerical scheme}
\label{sec:numerics}

Our numerical scheme for the fluid evolution 
is based on high resolution shock capturing methods (HRSC)
that have been used extensively in recent years.
They are described in details in numerous papers and review articles \cite{rjlbook,rjl98,ibanez,romero}.
Therefore we are going to describe only those features of our code that are 
new or not completely obvious.
All the calculations have been performed with quadruple precision arithmetic
in order to tune the critical parameter $p$ to as many as $30$ significant digits.

We used adaptive mesh refinement (AMR) in order to follow the evolution
on an ever decreasing spatial and time scales.
Another feature we implemented was a ``true'' vacuum, i.e.,  
in the vacuum regions we do not keep a small residual fluid atmosphere
(a so called {\em floor}) but the fluid variables are set to zero values.

The regularity conditions for fluid variables are achieved by using 
{\em reflective} boundary conditions at the origin $r=0$.
We use {\em outflow} boundary conditions at the outer boundary
(although there should be no fluid crossing the outer boundary).

To solve for the geometry we first solve for $a$ using the equation 
\eref{eq2.27}.
We demand that $a(r=0)=1$ and integrate the equation outward.
$\alpha$ is obtained from equation \eref{eq2.26} by integrating it inward.
It is initialized by setting $\alpha(r=r_{\rm max}) = -1/a(r=r_{\rm max})$.

\subsection{Implementation of AMR}

Our implementation of AMR can be viewed as
an evolution on a dynamically changing  non-uniform grid 
with hierarchical structure\footnote{In this context a hierarchical grid
is a grid in which the size of two neighboring cells may differ at most by
the refinement ratio (in our case $2$).}.
We use a single global time step that is determined by the size of 
the smallest cell.

The grid structure changes if certain criteria are not met.
The tests are performed at every fixed number of steps (at a {\em checkpoint}).
The criteria we first implemented were based on the truncation error
estimate.
We compare a one step update on a $2$:$1$ coarsened grid
with two step update on the original grid. 
The problem with this approach was that occasionally we experienced problems
with the update during the coarse grid step.
This certainly could be fixed but we opted for a simpler criteria.
In essence we want to guarantee that all the features of the solutions
are well resolved.
We demand higher resolution in regions with larger differences in gradients
of the variables.
The cells that do not meet our criteria are flagged (we can use more than one criterion).

After flagging we apply {\em buffering}, i.e.,
we flag nearby regions around the originally flagged cells.
The size of the buffer regions is chosen so that each of the
flagged cells could affect at most the cells within the buffer region
during the evolution until the next checkpoint.
Or, equivalently, we can say that only information from the buffer region
could have reached the flagged cell during evolution from the previous checkpoint.

After the buffering step the grid is adjusted so that all the flagged cells are
refined (we use $2$:$1$ refinement ratio). 
We demand that the refinement algorithm  preserve the hierarchical structure.

At this point we restart the evolution from the last {\em successful}
checkpoint but with the newly created grid.
All the variables are interpolated onto the new grid.
If no cells were flagged (the grid is not changed) then
the current state is stored (the grid structure and
all the variables) and the evolution continues.

In systems where discontinuities may be present we must be a little bit more careful
with the refinement scheme. 
Typically, our criteria would always fail at discontinuities and therefore
a naive application of the aforementioned rules would lead to an uncontrolled 
refinement (in practice, we always set a maximum refinement level).
This is not desirable since this would slow down the evolution tremendously.
Moreover, our scheme is ``shock capturing'' therefore it should treat 
discontinuous solutions properly.
In general this issue is not trivial an some clever scheme must be applied to
deal with this.
Our situation is slightly easier since we qualitatively know the dynamics.
Although we do not expect shocks in the critical solutions we do have shocks
in the computational domain.
These are present in the ``outer domain'', i.e., beyond the central self similar
 region.
We therefore employ a simple approach --- we introduce different refinement 
maxima for the inner and outer regions (the outer region maximum is much smaller
 than the inner one).

\subsection{Treatment of the vacuum}

Vacuum regions always pose a problem in numerical hydrodynamics.
Technically, the problem is that in very rarefied regions one part of the
numerical scheme fails --- in particular the conversion from conservative 
to primitive variables.
It simply happens that there are no physical values of the primitive variables
that correspond to the updated conservative variables.
These regions are, however, dynamically completely unimportant since the energy and
momentum densities there are negligible.

A standard approach is to maintain a minimal ``atmosphere''
of fluid everywhere with the density several orders of magnitudes smaller
than the typical densities in the system.
Although we believe that there is nothing wrong with this approach and it
would most certainly work in our situation we tried to develop a different approach.

One nice feature of a ``true'' vacuum is that there are no
fluxes through the outer boundary and therefore no artificial reflections.

Of course we can not really solve the problem because of the
non-existence of the exact Riemann solution at the vacuum boundary region.
What we propose is a set of ideas which typically work in practice.
They are not universal and one might need to adjust them slightly for different
systems.
Sometimes not all of them need to be used.
In general, the situation is more complicated with stiffer fluids ($\Gamma\to2$)
but we were able to use the scheme even for critical collapse of an ideal fluid
with $\Gamma=2$.
We also used this approach in 2D modeling of axisymmetric fluid accretion 
onto black holes and some preliminary 3D runs of neutron stars.
For the type of systems studied in this paper ($\Gamma\to 1$) the scheme
works in its simplest form.

The basic ideas are as follows.
\begin{enumerate}
\item 
  set the flux between two vacuum cells to zero
\item
\label{it2}
  the flux between vacuum and non-vacuum cell is
  calculated using the Tadmor's scheme (as described in \cite{Tadmor})
\item
  after the update of the conservative 
  variables set to vacuum all cells with fluid levels below certain threshold
\end{enumerate}

The purpose of item \eref{it2} is to use a simple and robust update scheme
which relies only on the characteristic speeds and not on the 
full spectral decomposition typically needed in Roe type of solvers.

Often, the approach described above is sufficient.
This was the case for all the calculations we performed for purpose of
this paper.
Since we used quadruple precision and our fluid is extremely soft
we did not even need to apply the rule number 2.

Sometimes we run into problems even after applying the above ideas.
In this case we should always try to adjust the vacuum threshold up to
the maximum level we can afford without loosing a significant amount of 
fluid in the system.

If we use AMR and we run into trouble at some cell we can flag the cell 
and invoke an ``emergency'' refinement and restart, especially if the cell
 is not at the vacuum boundary.
We used this approach when we tested the numerical scheme on
$\Gamma=2$ ideal fluid.
If the difficulty arises at the vacuum boundary (which happens in the
 majority of cases) we can simply set the cell to the vacuum values.
 
In general,  it is likely that the velocities near the vacuum boundary will not 
be smooth and will be be highly relativistic but this should not have
any adverse effect on the dynamics (as long as the scheme works).

\section{Conclusions}
\label{sec:conclusion}

In this paper we calculated the critical solutions and scaling exponents for
values of $\Gamma$ close to $1$ using both the CSS ansatz and by numerically solving the
full set of equations for spherically symmetric ultrarelativistic fluid collapse.
Moreover, by using the CSS ansatz we obtained and solved the equations 
for the limiting case $\Gamma=1$.
In the limit of $\Gamma\to 1$ the general relativistic critical solution converges to
the Newtonian Hunters type (a) solution.
The limiting value of the scaling exponent
\be
  \lim_{k\to 0} \gamma(k) = 0.1056669768
\n{eq7.14}
\ee
can be therefore interpreted as the scaling exponent for critical collapse 
in Newtonian gravity.
This is consistent with conclusions drawn from previous work \cite{Ori_Piran_90,Harada_Maeda_2001}.

We also addressed the possibility of generic naked singularity formation in the 
supercritical regime.
Our calculations support the fact that the supercritical collapsing solutions for $\Gamma-1\le0.01$
converge to the GRLP solution (an in the limit of $\Gamma\to 1$ to the Newtonian
Larson-Penston solution) and from \cite{Ori_Piran_90} then follows that these solutions
contain a naked singularity.
In other words, the GRLP solution is an endstate for a supercritical collapse for a slightly 
supercritical dataset for our family of initial 
data\footnote{The values of the parameter $p$ for the supercritical calculations
 were taken to be $0.1$\% larger than $p_*$ for the $\Gamma-1=10^{-6}$ runs and $4$\% larger than
$p_*$ for the $\Gamma-1=10^{-2}$ runs. These values were chosen randomly and were not tuned in any way.}.
It would be interesting to further explore the structure of the solution space, in particular, to 
search for a supercritical collapse with black hole as an endstate.

The numerically found solutions and scaling exponents agree very well 
(to a fraction of a percent) with those obtained by using the CSS ansatz.
To numerically solve the equations we used advanced numerical methods
including AMR, special vacuum treatment and quadruple precision floating point
arithmetic.

\section*{Acknowledgments}
This research was supported by 
the Natural Sciences and Engineering Research Council of Canada (NSERC)~\cite{NSERC},
the Canadian Institute for Advanced Research (CIAR)~\cite{CIAR},
the Alberta Science and Research Authority (ASRA)~\cite{ASRA},
the British Columbia Knowledge Development Fund (BCKDF)~\cite{BCKDF},
and the University of British Columbia. 
In particular, some of the calculations described herein were run on 
CFI/ASRA/BCKDF funded hardware: the WestGrid cluster 
{\tt glacier.westgrid.ca}~\cite{WestGrid} and the UBC PHAS/MECH cluster
{\tt vnp4.physics.ubc.ca}~\cite{vnp4}.
The author would like to thank Matt Choptuik, William Unruh and Carsten Gundlach
 for stimulating discussions and support of this work.
%

\end{document}